% pahresssproceedings.tex

% Invited talk by M. A. Pahre (Caltech) at the
%	Second Stromlo Symposium
%	22-26 August 1996
%	Canberra, Australia
% 
% Bibliographical information for this paper:
%	Pahre, M. A., \& Djorgovski, S. G.  1997, in The Nature of Elliptical
%	  Galaxies, ASP Conf. Ser. Vol. ??, eds. M. Arnaboldi, G. S. Da Costa,
%	  \& P. Saha, in press
% 
% Contact address:
%	map@astro.caltech.edu
%	Caltech, MS 105-24
%	Pasadena, CA  91125
%	USA
%	Office:  +01 (818) 395-4001
%	FAX:     +01 (818) 568-9352

% Put definitions at top (before inputting Stromlo ``style'' file) so that 
%	they can be overridden (if the editors prefer a different format)
\def\etal{{\sl et al.}}
\def\apj{ApJ}
\def\apjs{ApJS}
\def\apjl{ApJ}
\def\mnras{MNRAS}
\def\pasp{PASP}

\def\aa{A\&A}

\input stromlo

\title Observational Constraints on the Origins of the Fundamental Plane

\shorttitle Fundamental Plane

\author M. A. Pahre \& S. G. Djorgovski

\shortauthor Pahre \& Djorgovski

\affil Palomar Observatory, Caltech, USA

\abstract We review the concept of the Fundamental Plane (FP), and present 
new results on the near--infrared FP.
We show that the $K$--band FP differs both from the optical form and the 
virial expectation under the assumption of constant $(M/L)$ and homology.
Systematic variations of only the stellar populations parameters (age, 
metallicity, IMF) cannot reproduce the slopes of the FP 
simultaneously at both the optical and near--infrared wavelengths.
There appears to be an additional effect which could be due to a systematic 
departure of the velocity distributions of elliptical galaxies from a 
homologous family.  
In order to distinguish between the different possible stellar 
populations parameters, the FP and its projections can be observed at a range 
of redshifts.
We describe several such evolutionary tests:  in the intercept of the
color--magnitude relation; in the intercept of the Kormendy surface 
brightness--radius correlation; and in the intercept of the FP itself.
All three tests are fully consistent with the population of cluster elliptical
galaxies having formed at high redshift; this contradicts large,
systematic age variations as a significant contributor to the slope of the FP.

\section What is the Fundamental Plane?

The Fundamental Plane (FP) is a bivariate family of correlations of the 
properties of elliptical galaxies (Djorgovski \& Davis 1987; 
Dressler \etal\ 1987).
The properties that are usually correlated for the FP are the
half--light radius $r_e$, the mean surface brightness (SB) 
$\langle\mu\rangle_e$ interior to that radius, and the central velocity 
dispersion $\sigma_0$.
Other parameters can be used in the FP, such as the substitution of a color 
or line index for $\sigma_0$ (de Carvalho \& Djorgovski 1989), 
or the substitution of luminosity for radius.
Some monovariate correlations also represent important properties among 
ellipticals, such as the distance--independent Mg$_2$--$\sigma$ relation 
which may be a good age indicator (see Bender, this volume).

The FP can be projected onto any two axes out of the three variables.  
Examples of these projections are the color--magnitude relation, the 
Kormendy radius--SB relation (Kormendy 1977), and the Faber \& Jackson (1976) 
relation between luminosity and velocity dispersion.
The $D_n$--$\sigma$ relation is another example, as it was constructed as a 
nearly edge--on projection of the FP (Dressler \etal\ 1987); its residuals 
should correlate slightly with SB (Gunn 1988), as was observed by Lucey, 
Bower, \& Ellis (1991a).

The FP is usually expressed as a correlation of the observed properties of
ellipticals, but the correlations of real interest are those of the
underlying physical properties, such as galaxy mass, mass distribution,
velocity distribution, luminosity density, and stellar populations parameters.
The key to understanding the significance of the existence of a FP lies in 
determining the how the observed correlations translate into constraints on
underlying physical properties.
For example, stellar populations effects can be found in the line strengths,
colors, luminosity, and surface brightness terms which enter into the FP.

The FP is not necessarily a universal correlation for all environments.
A systematic difference may exist between field and cluster ellipticals in 
both the intercept and scatter of the FP (de Carvalho \& Djorgovski 1992).
There may be slight differences between the slopes of the FP in the Coma and 
Virgo clusters (Lucey \etal\ 1991a) and between the cores and halos of several 
rich clusters (Djorgovski, de Carvalho, \& Han 1989; Lucey \etal\ 1991a).
Finally, there are also signs of systematic differences (possibly due to 
environment) between ellipticals in the Coma cluster and the Hydra--Centaurus 
supercluster (Guzman 1995).  
For the remainder of this paper, we will concentrate on ellipticals found 
in rich clusters of galaxies, as these seem to show the greatest homogeneity 
of their properties.

\section Constraints Placed on Ellipticals by the FP

Elliptical galaxies in the Coma cluster have the following FP in the 
$V$--band (Lucey \etal\ 1991b):
$$r_e \propto \sigma_0^{1.23} \langle\Sigma\rangle_e^{-0.82} . $$
If we assume both that light traces mass (i.e. all galaxies have the same 
mass--to--light ratio) and that elliptical galaxies form a homologous family, 
then application of the virial theorem predicts the FP relation to be:
$$r_e \propto \sigma_0^{2} \langle\Sigma\rangle_e^{-1} . $$
The deviation of the observed $\langle\Sigma\rangle_e$ exponent from the 
virial expectation appears to be roughly independent of wavelength 
(compare Lucey \etal\ 1991b at $V$ with Pahre, Djorgovski, \& de Carvalho 
1995 at $K$);  it will not be discussed further here, but this deviation 
may partly be due to structural departures from a homology 
(Graham \& Colless, this volume).  

The deviation of the $\sigma_0$ exponent from the virial expectation
could be caused by a breakdown of either of the two assumptions.
A systematic variation in mass--to--light ratio along the FP could 
be due to variations in the stellar content (age, metallicity, or IMF) 
among ellipticals.
The resulting dependence is then $(M/L) \propto M^\alpha$, where 
$\alpha \sim 0.24$ in the $V$--band.  
The value of $\alpha$ should also change with wavelength if it is due to
stellar populations effects.
Systematic variations in the amount of dark matter or its distribution could
also create a variation in $(M/L)$ along the FP.

If the $\sigma_0$ exponent is caused by systematic departures of the structure
and dynamics of elliptical galaxies from a homology, then this effect should 
be strictly independent of wavelength.
There are reasons to suspect that structural nonhomology is present among
elliptical galaxies due to variations in the shapes of the light profiles
(Burkert 1993; Caon, Capaccioli, \& D'Onofrio 1993; Hjorth \& Madsen 1995;
Graham \& Colless, this volume).
A direct analysis of the effects of structural nonhomology on the FP,
however, suggests that it is insufficient to explain the slope for the
$\sigma_0$ exponent in the $V$--band (Graham \& Colless, this volume).
Dynamical nonhomology is motivated in numerical simulations of 
dissipationless merging (Capelato, de Carvalho, \& Carlberg 1995), but a 
wide--variety of analytical models have yet to be explored for this effect.

Regardless of explanation for the slopes of the FP, another constraint due to 
the FP is its extremely small scatter (see Renzini \& Ciotti 1993).
It is still unclear if the thickness of the FP has ever been resolved,
or if its thickness can be entirely explained by observational errors.
Line indices from high--quality spectra of nearby ellipticals (dominated by 
field and Virgo galaxies), when compared to the stellar populations models of 
Worthey (1994), suggest that there could be a complicated interplay between 
age and metallicity that manages to keep the FP thin (Worthey, Trager, \& 
Faber 1996).  
The result of those data and model comparisons is a prediction that 
$(M/L)_K \propto {\rm constant}$.  

\section The Near--Infrared Fundamental Plane

As seen in \S 2, the slope of the near--infrared FP is a direct test on
a number of models for the origin of the FP:  the wavelength dependence of
stellar populations effects (age, metallicity, or IMF, taken separately);
nonhomology; and the age--metallicity correlations of Worthey \etal\ (1996).
For this reason, we undertook a large, $K$--band, wide--field imaging survey
of $> 200$ elliptical galaxies in 9 rich clusters and 2 loose groups.
The data for this project were obtained with a IR cameras on the Palomar
60'' telescope and the Las Campanas 40'' and 100'' telescopes.
Velocity dispersion data were drawn from the literature.
The data have all been taken and reduced; the analysis is currently underway,
and preliminary results are described below.  
An analysis of $1/3$ of the dataset in five clusters were presented by 
Pahre \etal\ (1995).  

% Figure 1.  Multiple panel FP for individual clusters.

\figureps[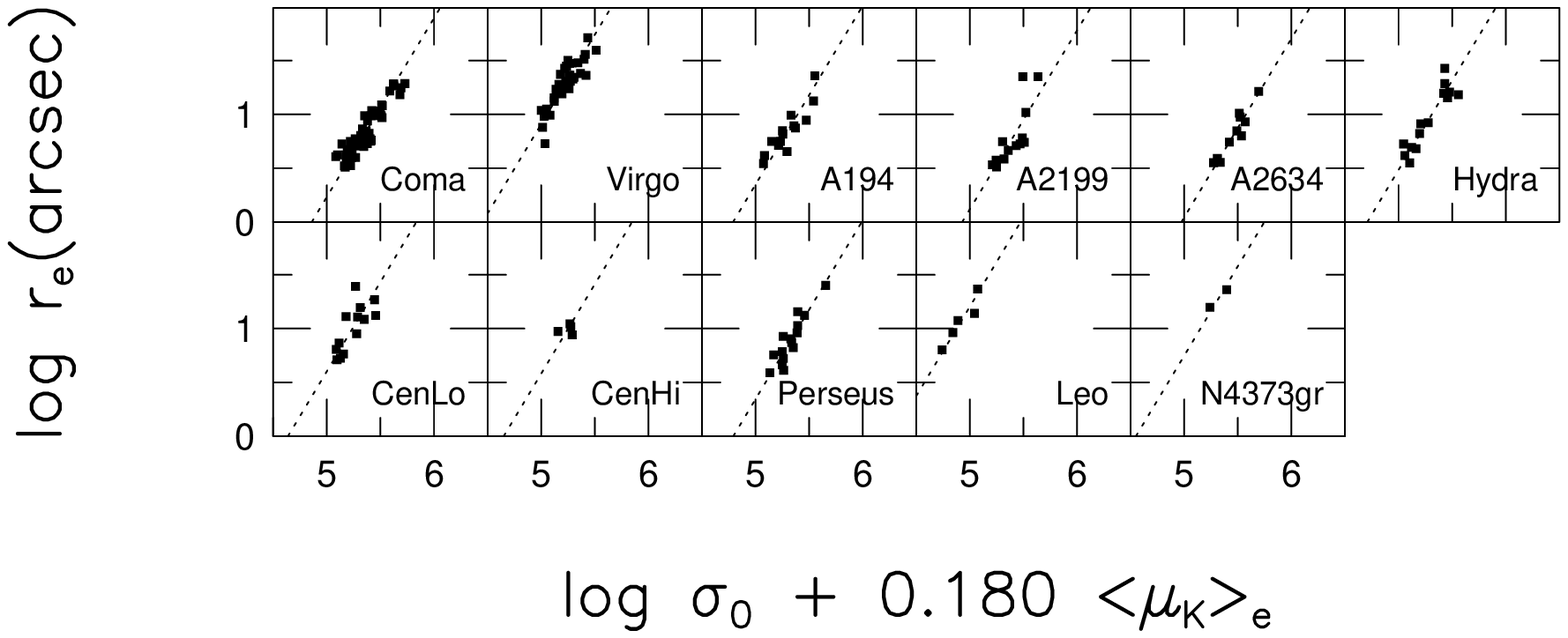,0.9\hsize] 1. The near--infrared Fundamental Plane.
The dotted lines displayed correspond to the least--squares fit (i.e. Eq.~3) 
in which the slope is fixed for all clusters, but the intercepts are allowed 
to vary for each cluster.  
A common FP appears to represent the data in all clusters well.

We find that the FP in the $K$--band is well--represented by the relation
$$r_e \propto \sigma_0^{1.66 \pm 0.09} 
	\langle\Sigma\rangle_e^{-0.75 \pm 0.03} . $$
It is immediately apparent, via comparison with Eq.~1, that the
near--infrared FP has a larger exponent in $\sigma_0$ than in the $V$--band;
by comparing with Eq.~2, it is also clear that the near--infrared FP
still deviates from the virial expectation at high confidence.
The slope of the FP varies somewhat between clusters, but this effect is small
and is being investigated.  
Nonetheless, common FP appears to represent all clusters quite well, 
as can be seen in Figure 1.
We find the identical effect as in Guzman (1995) for a systematic 
difference between the Coma cluster and the Hydra--Centaurus supercluster 
in the distance--independent relationship between $(D_K - D_V)$ and $\sigma_0$.
We note that the FP we derive in the $K$--band is fully consistent with that 
from photoelectric photometry (Recillas--Cruz \etal\ 1990, 1991; 
Djorgovski \& Santiago 1993).

\section An Explanation for the Slope of the FP

Now that the FP has been derived with imaging data at both optical and infrared
wavelengths, it is possible to construct and test models that might explain the
slope of the FP.
A comparison of the expected wavelength effects of various stellar populations
models (Worthey 1994), and nonhomology models, is demonstrated in Figure 2.
The extent of the FP was assumed to be $4$~mag in luminosity, and the ordinate
is the power law in the relation $(M/L) \propto L^\beta$ (where the SB term is
near unity and hence ignored).
It is clear from the diagram that no stellar populations models by themselves
are able to fit the slope of the FP at all wavelengths; some other effect is
required which appears to be rougly independent of wavelength.  
Likewise, models of nonhomology alone cannot explain the data.
Instead, some combination of the two effects is needed, as can be seen by 
the rough agreement in the right--hand panel of Figure 2.

% Figure 2.  
%	Left panel:	Wavelength vs. Beta, data points + stellar pops models
%	Middle panel:	Wavelength vs. Beta, data points + nonhomology models
%	Right panel:	Wavelength vs. Beta, data points + stellar pops 
%     						+ nonhomology models

\figureps[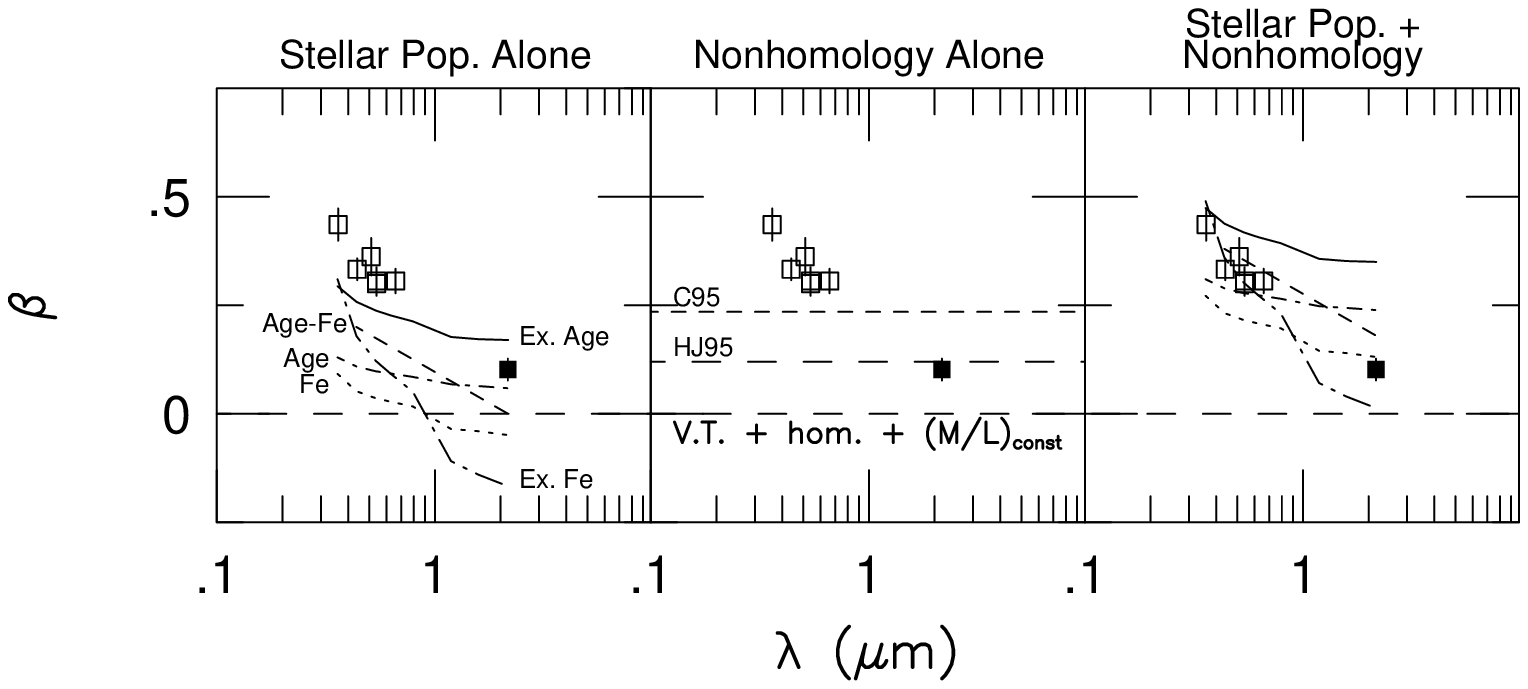,0.9\hsize] 2.  Fitting stellar populations models and
deviations from homology to the slope of the FP at different wavelengths.
The data are for cluster ellipticals at $U,B,g,r$ from J\o rgensen, Franx, 
\& Kj\ae rgaard (1996), $V$ from Lucey \etal\ (1991a), and $K$ (this paper, 
solid symbol).
The stellar populations models [left] are labelled as follows:  
$-0.1 < [Fe/H] < +0.1$ (Fe); $-0.5 < [Fe/H] < +0.25$ (Ex. Fe); 
$9 < t < 14$~Gyr (Age); $3 < t < 9$~Gyr (Ex. Age); 
and the age--metallicity model of Worthey \etal\ (1996; Age--Fe).
Not even the extreme models of age or metallicity can explain the slope of 
the FP at all wavelengths.
The nonhomology models [middle] are from Capelato \etal\ (1995; C95) and 
Hjorth \& Madsen (1995; HJ95), and also cannot explain the data.
Adding the effects of nonhomology to those of the stellar populations models 
[left], however, allows room for significant agreement between the models 
and the slope of the FP at the different wavelengths.

Some of the extreme stellar populations models in Figure 2 can be excluded 
on the basis of other information.
The extreme age model has a larger age spread (factor of 3) than even the 
models of Worthey \etal\ (1996; a factor of two), and the extreme metallicity 
model would predict a slope for the color--magnitude relation that is much 
steeper than observed (Bower, Lucey, \& Ellis 1992).  
Hence, modest stellar populations effects are favored by this analysis.

It is common throughout the literature to refer to the deviation of the slope 
of the FP from the virial expectation as constituting an ``observed dependence
of mass--to--light ratio on luminosity.''  
As the above argument suggests, however, this might in part be due to 
deviations of the velocity distributions of ellipticals from a homologous 
family, and not due to intrinsic variations in $M/L$.
Hence the term ``observed dependence of $M/L$ on $L$'' would be, at least in 
part, a misnomer.

The composite model described above makes several predictions.
First, the slope of the FP should increase with increasing aperture used to 
measure the velocity dispersion (i.e. Capelato \etal\ 1995).  
While it is generally known that velocity dispersions decrease with aperture, 
it is not clear if this is a systematic effect, i.e. that $\sigma$ decreases 
more rapidly with $(r/r_e)$ for the larger ellipticals.
There is such a hint in the analysis of J\o rgensen \etal\ (1995) or the data
of D'Onofrio \etal\ (this volume), but the scale of the observed effect may 
not be as dramatic as in the numerical simulations of Capelato \etal\ (1995).
More work is needed to address this point.
Second, if there are systematic deviations from a homology, then the stellar 
populations contributions to the slope of the FP are smaller than previously 
thought; thus there should be less evolution in the slope of the FP with 
redshift.

\section The FP and Its Projections at High Redshift

If there exists a significant, systematic spread in ages among the cluster
elliptical galaxy population, then an excellent test of this is to observe 
the FP and its projections at higher redshift (see also Franx, this volume).

\subsection Color Evolution of the Early--Type Population in Clusters

% Figure 3:  Evolution in $(U-V)_0$

\figureps[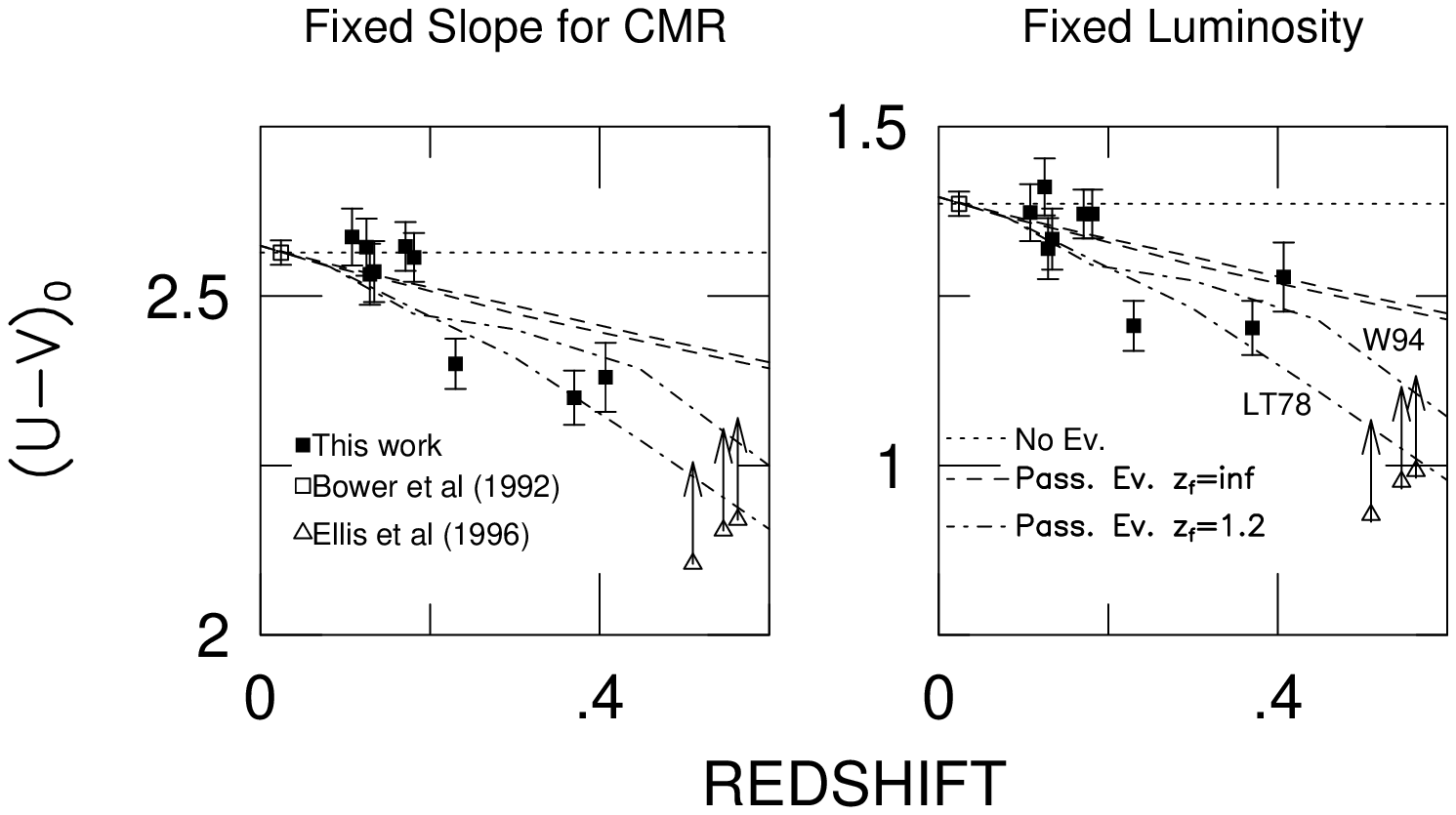,0.7\hsize] 3.  Color evolution in restframe 
$(U-V)$ for the early--type population in rich clusters of galaxies.  
A $k$--correction appropriate to the observed system ($U,V$ at $z<0.15$, 
$B,R_{\rm C}$ at $0.15 < z < 0.45$, and $V,I_{\rm C}$ at $z>0.5$) has been 
applied to transform into the restframe $U_0,V_0$.
The left panel is the intercept at $V=0$ assuming a constant slope for the CMR 
of $-0.08$, while the right panel is a measure of the color of the CMR at a 
fixed luminosity of $M_V = -20.185$ ($H_0=75$, $\Omega_0=0.2$, $\Lambda_0=0$).
There is a bluing trend with redshift for $z > 0.2$.
Data at $z\sim 0.5$ from Ellis \etal\ (1996) are included, although those 
authors note the substantial absolute calibration uncertainty of their data 
(as indicated by the arrows).
The models are from Worthey (1994; W94) and Larson \& Tinsley (1978; LT78).

The first step in observing the FP at high redshift is to define a sample
of galaxies in a given cluster.  
We have developed a quantitative method by which to identify the cluster 
elliptical population in a complete, reliable, and robust way using multicolor 
CCD imaging data (Pahre 1996, in preparation).
The key ingredient to this method is the use of a second color to discriminate
low--redshift cluster ellipticals from high--redshift field spirals;
this allows the individual color constraints to be applied weakly 
(i.e.~allowing a scatter of $0.2$~mag to $0.4$~mag in each color).
A morphological cut from the concentration index (Abraham \etal\ 1994) is
also included.
These selection criteria can be tested using Monte Carlo simulations
to determine if the final results are biased in any way.

This method immediately produces a measurement of the color evolution of the
early--type population in rich clusters using the intercept of the 
color--magnitude relation (CMR).
By using standard filters that progressively track the restframe $(U-V)$ 
with redshift, the uncertainty due to $k$--corrections can be reduced.
The results for the first nine clusters we have studied at $0.08 < z < 0.41$
are displayed in Figure 3.  
There is a clear bluing trend with redshift for $0.2 < z < 0.41$ which 
indicates a high redshift of formation for the stellar content of the 
cluster elliptical galaxy population as a whole.
If there were a conspiracy such that at all redshifts the reddest galaxies 
were always, say, 3~Gyr old, then no bluing trend would be seen; this is 
inconsistent with the data.
The small scatter of the color--magnitude relation at $z \sim 0.5$ 
(Ellis \etal\ 1996) could be explained by either a highly--synchronous 
formation of ellipticals at modest redshift, or by a less--synchronized 
formation at high redshift.
The color evolution in Figure 3 is a separate constraint which can 
discriminate between these two alternatives: the latter explanation appears 
favored.
Accurate color measurements at higher redshifts using similar selection 
criteria are clearly needed to strengthen and extend this result.
Our results are similar to those of Rakos \& Schombert (1995), although we 
find that the bluing trend begins at a lower redshift.

\subsection Evolution in the Kormendy Radius--Surface Brightness Relation

The Kormendy (1977) relation between SB and radius is a useful projection of 
the FP because it does not require the measurement of velocity dispersions.  
Thus it can be constructed using ground--based imaging under good seeing 
conditions or from HST imaging.
We have done an exploratory study (Pahre, Djorgovski, \& de Carvalho 1996) 
using Keck $K$--band imaging for Abell~2390 ($z=0.23$) in $0.45''$ seeing and 
HST/WFPC--2 F702W data publicly available for Abell~851 ($z=0.407$).
For the latter cluster, the HST images were calibrated onto the ground--based 
$R_{\rm C}$ photometric system, then transformed to the $K$--band using colors 
obtained at the Palomar 60--inch telescope.  
The results show the power of this method in measuring the luminosity 
evolution (via the SB term) of the Kormendy relation for a fixed metric 
scale, and are displayed in Figure 4.  
The Tolman signal is clearly detected, as is a luminosity evolution of 
$0.36 \pm 0.14$~mag in the $K$--band out to $z=0.4$.  
This result is fully consistent with that found by Barrientos \etal\ (1996) 
and Schade \etal\ (1996).

% Figure 4:  Evolution in the Kormendy relation

\figureps[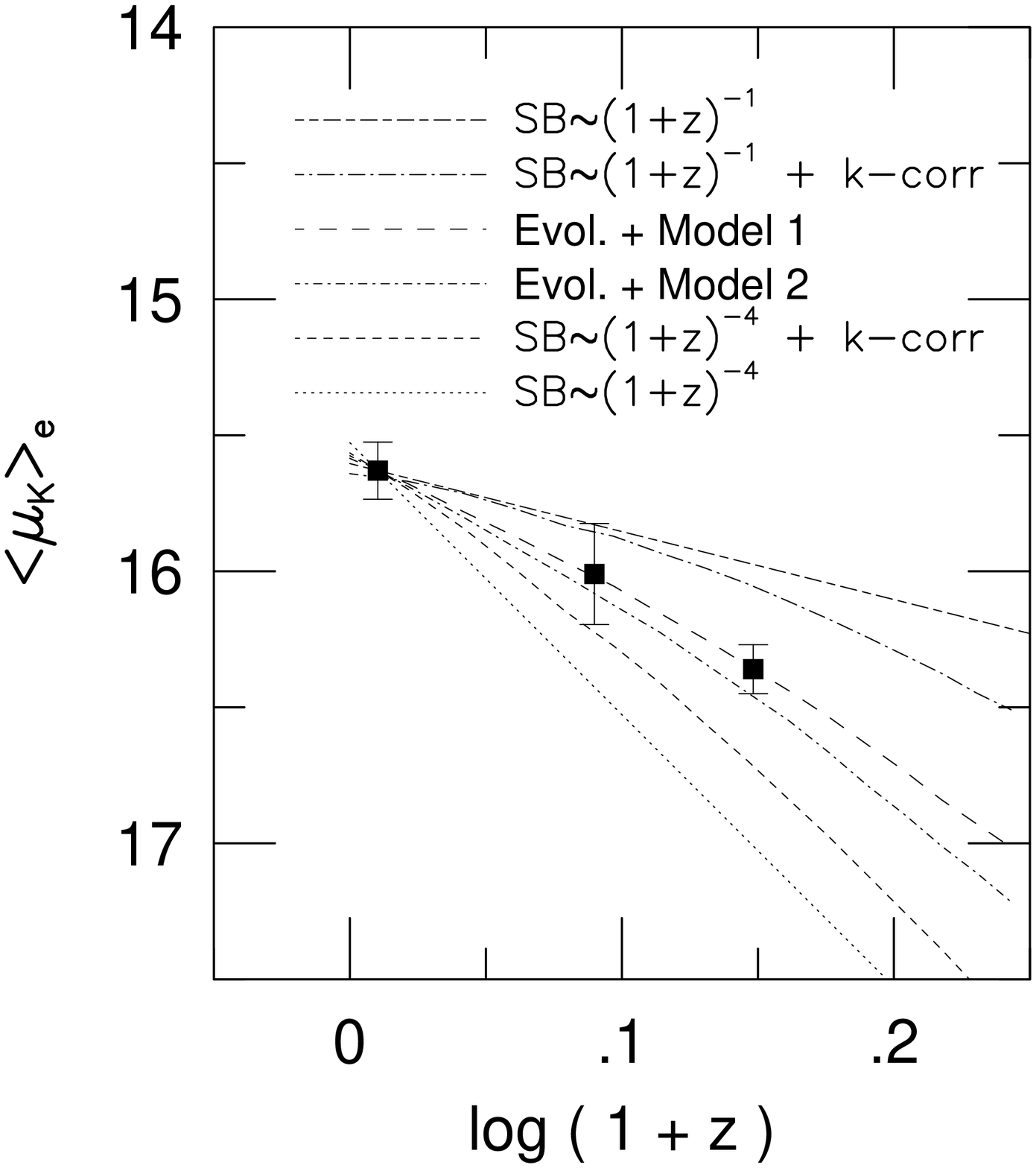,0.3\hsize] 4.  Evolution in the $K$--band Kormendy
radius--SB projection of the FP, with comparison to different cosmological 
models (see Pahre \etal\ 1996).  
A correction has been applied to the tired--light models (where SB dims as 
$(1+z)^{-1}$) in order to account for the different metrics in the 
non--expanding and expanding cosmologies used for evaluating the intercept 
at a fixed metric size.  
The data are consistent with an expanding universe in which the galaxies 
form at high redshift and evolve passively, and are inconsistent with a 
non--expanding cosmology.

\subsection The FP at High Redshift Using the Keck Telescope

The most challenging step towards constructing the full FP at high redshifts 
is the
measurement of a central velocity dispersion.
Early work by Franx (1993) required long integrations (9 hours) on the MMT in 
order to obtain the necessary S/N on a handful of galaxies at $z=0.18$; the 
Keck telescopes, on the other hand, are excellently--suited towards such 
observations.
Displayed in Figure 5 are six sample spectra (out of 19) taken in 1.4~hours 
in the same $z=0.18$ cluster which show the power of these telescopes in 
probing such redshifts.
It is now conceivable to pursue large surveys for the FP at high redshift in 
modest amounts of observing time.
Furthermore, small apertures can be used with the higher S/N Keck data, so 
that the the high redshift galaxies are measured in a similar way to low 
redshift galaxies.

% Figure 5:  Spectra of $z=0.23$ galaxies

\figureps[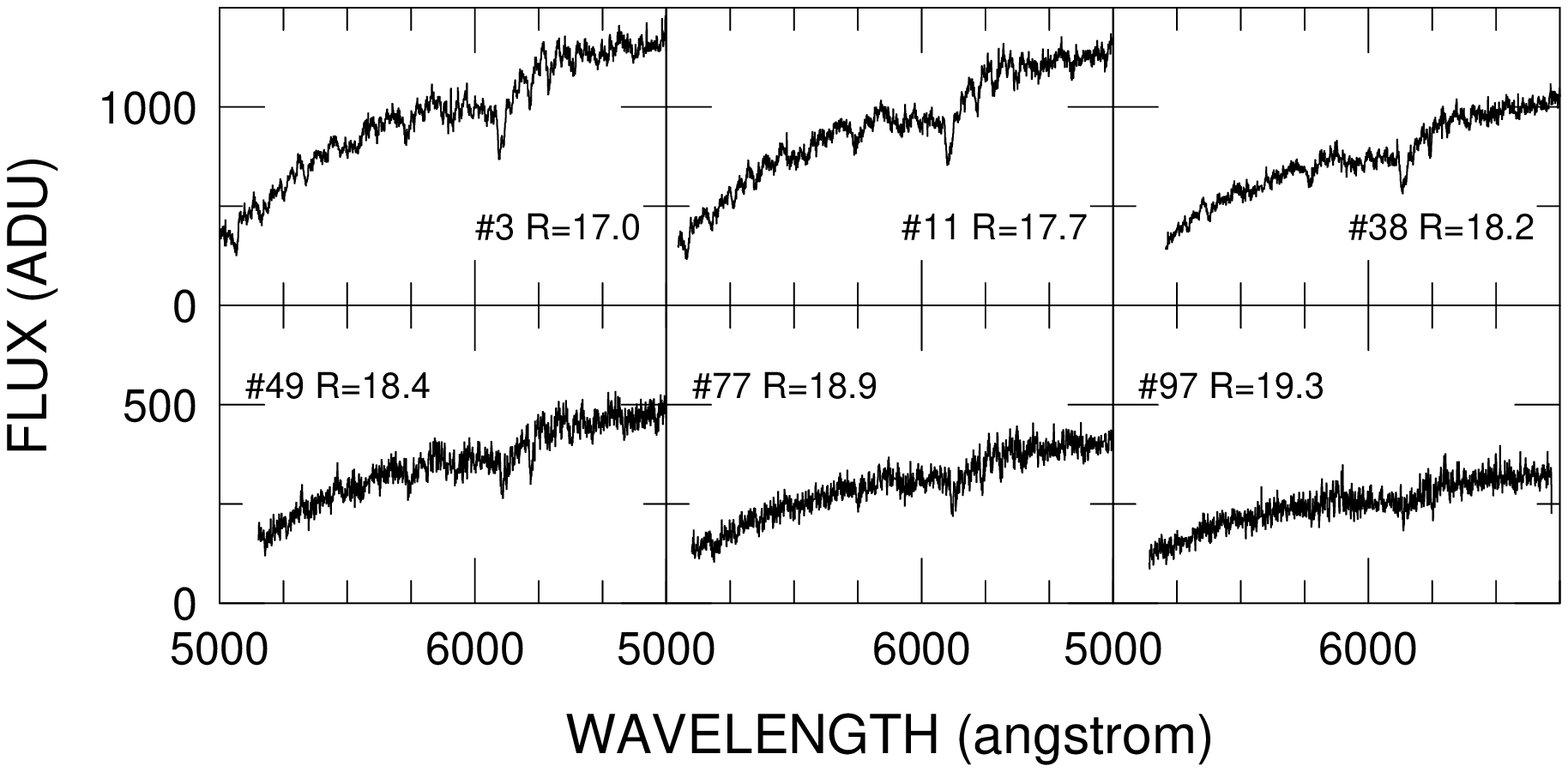,0.7\hsize] 
5.  Sample spectra of elliptical galaxies 
in Abell~665 ($z=0.18$) taken with the Keck~I telescope in a 1.4~hour 
integration.  
The galaxy identifications (where the number is the ranking in that cluster
among the early--type population) 
and the $R_{\rm C}$--band total magnitudes are given in each panel.  
The apertures measured here correspond to $0.9 \times 0.7$~arcsec$^2$, which 
allows direct comparison with low redshift galaxies independent of 
assumptions on aperture corrections.

We have begun a program to measure velocity dispersions in rich clusters of 
galaxies at $0.08 < z < 0.75$ on the Palomar 200--inch and the Keck telescopes.
Clusters at $z>0.15$ have been chosen from those with HST/WFPC--2 optical 
imaging publicly available in the archive.  
Surface photometry in the $K$--band is also measured on Keck in times of good 
seeing ($0.3$ to $0.5$~arcsec).
The preparatory two--color CCD imaging generates the galaxy lists through the 
selection criteria detailed in \S 5.1.
The first near--infrared FP at high redshift is displayed in Figure 6 for 
Abell~2390 ($z=0.23$), although we note that these data are still in the 
preliminary reductions stages.
As can also be seen from the paper by Franx (this volume), it is now clearly 
feasible to pursue large--scale surveys of the FP at these redshifts.
Passive luminosity evolution has been detected via the change in the intercept
of the FP by van Dokkum \& Franx (1996).

% Figure 6:  K-band FP

\figureps[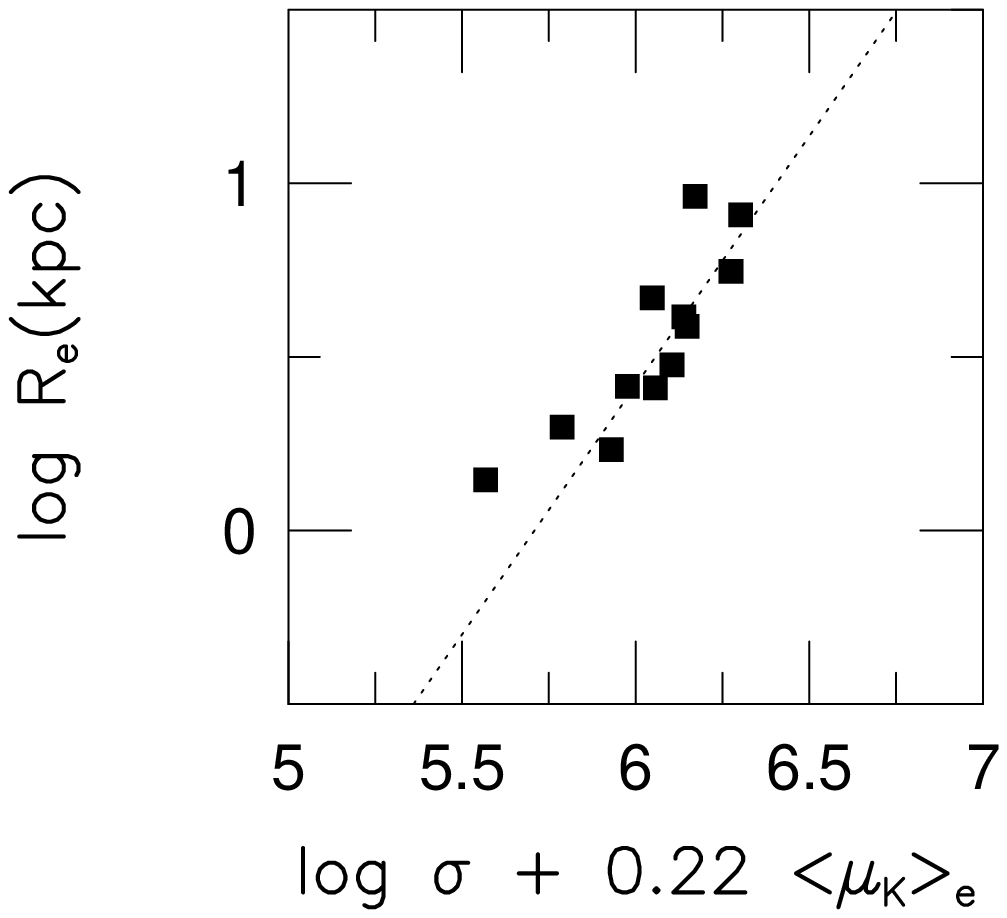,0.4\hsize] 
6.  The near--infrared FP for Abell~2390 
($z=0.23$) using $K$--band surface photometry and multiobject spectra from 
the Keck~I 10~m telescope.

There are a number of key questions that should be addressed by these surveys 
of the FP at high redshift.  
First, how does the intercept of the FP (or the SB term at fixed radius and 
velocity dispersion) vary with redshift?  
This will allow a more accurate Tolman SB test due 
to the reduced scatter of the FP itself; it will also allow the accurate 
measurement of the luminosity evolution of the early--type population.
Second, how does the slope of the FP vary with redshift?  
If the slope of the FP is produced by a systematic variation in age along it, 
then the slope of FP should also vary with redshift.
If the slope is due to metallicity variations and nonhomology, 
then the slope should change very little with redshift.
It will be necessary to obtain a large number of galaxies per cluster 
($\geq 20$) in order to measure the FP slope accurately.
Third, how do assumptions about aperture corrections on measurements of the 
central velocity distribution affect the slope of the FP?  
Fourth, how do the slopes of the distance--independent projections of the FP 
vary with redshift?  
The Mg--$\sigma$ correlation is a good age indicator (see Bender, this volume),
and so it should provide a good test of the contribution of age effects to 
the slope of the FP.
Fifth, does the slope of the FP correlate with other galaxy population 
parameters (like blue galaxy fraction, galaxy number density, or core profile 
shape or size)?
And finally, what do all of these results imply for the formation history
of the cluster elliptical galaxy population taken as a whole?

\section Summary

A number of different observational approaches have been described which 
provide constraints on possible origins for the FP.  
The wavelength variation of the slope of the FP has been shown to be 
inconsistent with all reasonable stellar populations models, requiring an 
additional effect such as departures of ellipticals from a homologous family.
The aperture effect on velocity dispersions is a strong additional 
constraint on dynamical nonhomology that will need to be explored further.
The evolution of the FP (Franx, this volume) and its projections (the 
color--magnitude and Kormendy relations) also suggest a high redshift of 
formation, which is in direct contradiction with possible large age spreads 
among ellipticals contributing to the slope of the FP.
Finally, the small scatter of the FP will continue to be an additional 
powerful constraint for all possible origins for the slope of the FP.

\acknowl MAP would like to thank the organizers for their hospitality and
their invitation to present this paper.
We would like to thank R. de Carvalho for many useful discussions and 
suggestions on a draft of this manuscript.
Travel to Australia was supported by a AAS travel grant to MAP.
This work was partially supported by NSF PYI award AST--9157412 to SGD, 
and the Bressler Foundation.

\references

% Abraham, R. B., Valdes, F., Yee, H. K. C., \& van den Bergh, S.  1994, 
% 	\apj, 432, 75
Abraham, R. B., \etal\  1994, 
	\apj, 432, 75

% Arag\' on--Salamanca, A., Ellis, R. s., Couch, W. J., \& Carter, D.  1993, 
%	\mnras, 262, 764
Arag\' on--Salamanca, A., \etal\  1993, \mnras, 262, 764

Arimoto, N., \& Yoshii, Y.  1987, \aa, 173, 23

Barrientos, L. F., Schade, D., \& L\'opez--Cruz, O.  1996, ApJ, 460, L89

% Bender, R., Burstein, D., \& Faber, S. M.  1992, \apj, 399, 462

% Bender, R., Ziegler, B., \& Bruzual, G.  1996, \apj, 463, L51

Bower, R. G., Lucey, J. R., \& Ellis, R. S.  1992, \mnras, 254, 601

Burkert, A.  1993, \aa, 278, 23

Caon, N., Capaccioli, M., \& D'Onofrio, M.  1993, \mnras, 265, 1013

Capelato, H. V., de Carvalho, R. R., \& Carlberg, R. G.  1995, \apj, 451, 525

de Carvalho, R. R.,  \& Djorgovski, S.  1989, \apjl, 341, L37

de Carvalho, R. R.,  \& Djorgovski, S.  1992, \apjl, 389, L49

% Ciotti, L., \& Lanzoni, B.  1996, \aa, submitted

Djorgovski, S., \& Davis, M.  1987, \apj, 313, 59

Djorgovski, S., de Carvalho, R., and Han, M.-S. 1988, in The Extragalactic 
	Distance Scale, ASP Conf. Ser. Vol. 4, eds. S. van den Bergh \& 
	C. J. Pritchett, 329

Djorgovski, S., \& Santiago, B. X.  1993, in proceedings of the ESO/EIPC 
	Workshop on Struc\-ture, Dy\-na\-mics, and Chem\-i\-cal Evo\-lu\-tion 
	of Early--Type Gal\-ax\-ies, ed. J. Danziger, \etal, ESO publication 
	No. 45, 59

% Dressler, A., Lynden-Bell, D., Burstein, D., Davies, R., Faber, S. M., 
% 	Terlevich, R. J., \& Wegner, G.  1987, \apj, 313, 42
Dressler, A., \etal\  1987, \apj, 313, 42

% Ellis, R. S., Smail, I., Dressler, A., Couch, W. J., Oemler, A., Butcher, H.,%	\& Sharples, R. M.  1996, \apj, submitted
Ellis, R. S., \etal\  1996, \apj, submitted

Faber, S. M., \& Jackson, R. E.  1976, ApJ, 204, 668

Franx, M.  1993, \pasp, 105, 1058

Gunn, J. E.  1988, in The Extragalactic Distance Scale, ASP Conf. Ser. Vol. 4, 
	eds. S. van den Bergh \& C. J. Pritchett, 344

Guzman, R.  1995, in proceedings of the Heron Island Workshop on Peculiar 
	Velocities in the Universe, {\+http://qso.lanl.gov/~heron/}

Hjorth, J., \& Madsen, J.  1995, \apj, 445, 55

J\o rgensen, I., Franx, M., \& Kj\ae rgaard, P.  1995, \mnras, 276, 1341

J\o rgensen, I., Franx, M., \& Kj\ae rgaard, P.  1996, \mnras, 280, 167

Kormendy, J.  1977, \apj, 218, 333

Larson, R. B., \& Tinsley, B. M.  1978, \apj, 219, 46

Lucey, J. R., Bower, R. G., \& Ellis, R. S.  1991a, \mnras, 249, 755

% Lucey, J. R., Guzm\'an, R., Carter, D., \& Terlevich, R. J. 1991b, \mnras, 
% 	253, 584
Lucey, J. R., \etal\  1991b, \mnras, 253, 584

% Murphy, D.~C., Persson, S.~E., Pahre, M.~A., Sivaramakrishnan, A., \& 
% 	Djorgovski, S.~G.  1995, \pasp, 107, 1234

Pahre, M.~A., Djorgovski, S.~G., \& de Carvalho, R.~R. 1995, ApJ, 453, L17

Pahre, M.~A., Djorgovski, S.~G., \& de Carvalho, R.~R. 1996, ApJ, 456, L79

% Persson, S. E., West, S. C., Carr, D. M., Sivaramakrishnan, A., 
% 	\& Murphy, D. C.  1992, \pasp, 104, 204

% Petrosian, V.  1976, \apj, 209, L1

Rakos, K. D., \& Schombert, J. M.  1995, \apj, 439, 47

% Recillas-Cruz, E., Carrasco, L., Serrano, P. G., \& Cruz-Gonz\'alez, I.  
%	1990, \aa, 229, 64
Recillas-Cruz, E., \etal\  1990, \aa, 229, 64

% Recillas-Cruz, E., Carrasco, L., Serrano, P. G., \& Cruz-Gonz\'alez, I.  
% 	1991, \aa, 249, 312
Recillas-Cruz, E., \etal\  1991, \aa, 249, 312

Renzini, A., \& Ciotti, L.  1993, \apjl, 416, L49

% Schade, D., Carlberg, R. G., Yee, H. K. C., \& L\'opez--Cruz, O.  1996, ApJ, 
%	464, L63
Schade, D., \etal\  1996, ApJ, 464, L63

% Stanford, S. A., Eisenhardt, P. R. M., \& Dickinson, M.  1995, \apj, 450, 512

van Dokkum, P. G., \& Franx, M.  1996, \mnras, 281, 985

Worthey, G.  1994, \apjs, 95, 107

Worthey, G., Trager, S. C., \& Faber, S. M.  1996, in Fresh Views of 
	Elliptical Galaxies, ASP Conf. Ser. Vol. 86, eds. A. Buzzoni 
	\& A. Renzini, 203

Zepf, S. E., \& Silk, J.  1996, \apj, 466, 114

% edit this out for the proceedings; edit in for MAP's personal version

\maintextmode
\parindent=\theparindent
\hang\noindent

\def\discussion{\medskip\maintextmode\bigskip\maybebreak{.1}
	\leftline{\bf Discussion}}
\def\Q#1\par{\smallskip\maintextmode\bigskip\maybebreak{.1}
	\leftline{\bf Question: #1}  }
\def\A#1\par{\smallskip\maintextmode\bigskip\maybebreak{.1}
	\leftline{\bf Answer: #1}  }

\discussion

\Q Mateo

Which method of measuring velocity dispersions at low S/N has worked
best?

\A Pahre

We have measured velocity dispersions with the fourier quotient, 
cross--correlation, and fourier fitting methods, but have not fully 
investigated this issue.  Our preliminary reductions suggest that there
is a small bias in the cross--correlation method, but not between the
other two.  The measurements in Figure 6 are from the fourier fitting
method.

\Q Silva 

Just a quick word of caution that current evolutionary 
population synthesis models do {\sl not} reproduce near-IR ($JHK$) colors of 
ellipticals.  So, using them to interpret the $K$--band FP is dangerous.

\A Pahre

Yes, such models do not reproduce absolute colors very well at all.  
Nonetheless, I have used them here solely as {\sl differential} 
models---changes in colors with redshift and $k$--corrections---in which 
many of those fundamental problems are substantially lessened.

\Q Dorman

(1)  Doesn't Worthey's model prediction of $(M/L)_K \propto {\rm constant}$ 
with age arises from the notion the $K$--band flux is dominated by giants 
that don't age strongly?  (2)  The $(V-K)$ color involves the ratio of 
turnoff flux to giant flux, where $K$ is sensitive to the latter and $V$ is 
sensitive to both.  Do you see $(V-K)$ evolution?  
(3)  If $(M/L)_K \propto L^\beta$, then what sort of model can make the 
giant flux change like that?

\A Pahre

(1)  Yes, I believe you are correct.  
(2)  This is something we intend to measure in the course of our current 
survey; Arag\' on--Salamanca \etal\ (1993) found evolution in $(V-K)_0$.  
(3)  The dependence is on the {\sl galaxy's} luminosity $L$ raised to the 
exponent $\beta$, hence there is no particular reason why the models proposed 
would not work.  The key is what galaxy formation process could produce 
such a dependence of stellar populations on galaxy luminosity.

\Q Yi

Fitting the observed $(U-V)$ colors as functions of redshift with (population 
synthesis) models should be carried out very carefully, because the model 
UV--U flux is ``extremely'' sensitive not only to age but also to several 
input assumptions, such as mass loss and horizontal branch mass distribution, 
which are often ignored in modelings.

\A Pahre

Certainly true.  Referring to Figure 3, I have included both the Larson \& 
Tinsley (1978) and Worthey (1994) models.  As you have noted previously, the 
former includes no HB and hence is too blue; the latter includes only a red 
clump for the HB and is probably too red.  The correct answer probably lies 
somewhere between the two.

\Q Zepf

The elemental abundance in cluster gas derived from ASCA observations require 
a very large number of massive stars to produce the high abundance of 
$\alpha$ elements.  The mass in this massive star population is sufficiently 
large that the remnants from this population are a significant fraction of 
the observed mass in ellipticals (roughly 30\%).  
Thus, there is a lot of room to get changes in $(M/L)$ with $L$ by varying 
the IMF with $L$.

\A Pahre

The models of Worthey (1994) do not include a galactic wind (such as in 
Arimoto \& Yoshii 1987) hence they cannot provide any dependance of $(M/L)$ 
on $L$ as such.  While your paper (Zepf \& Silk 1996) makes a case that 
varying the IMF could produce at least part of the variation of $(M/L)$ with 
$L$, it is still not clear if that would fully match the data in Figure 2.  
A wavelength--independent effect, such as nonhomology, might still be required.

% edit this out for MAP's personal version, as there is a \bye in the 
%	file questions.tex
% edit it back in for actual proceedings version

% \bye

\bye